\documentclass{article}

\usepackage{amsmath}
\usepackage{amsfonts}
\usepackage{amsthm}
\usepackage{graphicx}
\usepackage{comment}
\usepackage{algorithm, algcompatible}
\usepackage{bbm}
\usepackage{xfrac}
\usepackage{hyperref}
\usepackage{parskip}
\usepackage{cite}
\usepackage{url}
\usepackage[margin=1in]{geometry}
\usepackage{authblk}
\usepackage[symbol]{footmisc}

\title{Dynamic locational marginal emissions via implicit differentiation}

\author[1]{Lucas Fuentes Valenzuela}
\author[1]{Anthony Degleris\thanks{\textit{L. Fuentes Valenzuela and A. Degleris are co-first authors. Corresponding author: A. Degleris.}}}
\author[1]{Abbas El Gamal}
\author[2]{Marco Pavone}
\author[3]{Ram Rajagopal}

\affil[1]{Department of Electrical Engineering, Stanford University}
\affil[2]{Department of Aeronautics and Astronautics, Stanford University}
\affil[3]{Department of Civil Engineering, Stanford Engineering}

\usepackage[dvipsnames]{xcolor}
\usepackage{soul}

\newtheorem{theorem}{Theorem}

\newcommand{\reals}{\mathbf{R}}

\newcommand{\ones}{\mathbf{1}}

\begin{document}

\maketitle



\begin{abstract}
Locational marginal emissions rates (LMEs) estimate the rate of change in emissions due to a small change in demand in a transmission network, and are an important metric for assessing the impact of various energy policies or interventions. 
In this work, we develop a new method for computing the LMEs of an electricity system via implicit differentiation.
The method is model agnostic; it can compute LMEs for any convex optimization-based dispatch model, including some of the complex dispatch models employed by system operators in real electricity systems.
In particular, this method lets us derive LMEs for \textit{dynamic} dispatch models, i.e., models with temporal constraints such as ramping and storage.
Using real data from the U.S.\ electricity system, we validate the proposed method
against a state-of-the-art merit-order-based method and show that incorporating dynamic constraints improves model accuracy by 8.2\%.
Finally, we use simulations on a realistic 240-bus model of WECC to demonstrate the flexibility of the tool and the importance of incorporating dynamic constraints.
Namely, static LMEs and dynamic LMEs exhibit a normalized average RMS deviation of 28.40\%, implying dynamic constraints are essential to accurately modeling emissions rates.
\end{abstract}

\section{Introduction}
\label{sec:intro}

Policy-makers interested in decarbonizing the electricity grid require reliable emissions data in order to quantify the impact of a particular policy or intervention strategy.
Similarly, grid operators conducting generation and transmission expansion studies~\cite{Mo1991StochasticProgramming, Alguacil2003TransmissionApproach, Garcia-Bertrand2017DynamicPlanning, Majidi-Qadikolai2018Optimization-basedStudies} are increasingly looking to reduce emissions~\cite{Park2015StochasticEmissions, Bent2012GridReduction, Asgharian2017APlanning, Moreira2021Climate-AwareApproach, Degleris2021Emissions-awareDifferentiation} and require detailed information on the relationship between demand and emissions.
The need for this data will only grow as various systems, such as transportation and heating, begin to electrify and place additional strain on the grid.
Unfortunately, electricity systems depend on complex, interconnected transmission network structures with numerous operational constraints, making it difficult to attribute emissions to energy consumption at a particular time and place~\cite{deChalendar2019TrackingSystem}.

\textit{Emissions rates} are important metrics that quantify the amount of pollutant emissions, e.g., $\text{CO}_2$, $\text{SO}_2$, or $\text{NO}_\text{x}$, attributable to the consumption of energy. 
Researchers and decision-makers often examine \textit{average emissions rates} \cite{
Kirschen1997ContributionsFlows, 
Yu2019ResearchFlow, 
Kang2012CarbonNetworks, 
Kang2015CarbonModel, 
Li2013CarbonObligation
} 
which measure the average rate of emissions per MWh of energy consumed, and 
\textit{marginal emissions rates}~\cite{Hawkes2010EstimatingSystems, SchramOnFactors, MandelWattTimePrimer, McCormickMarginalWatttime, CorradiAAction} (also known as marginal emission factors or marginal emission intensities), which measure the rate at which emissions increase or decrease given a marginal change in energy demand. While average emissions rates quantify emissions over long periods, marginal emissions rates better quantify the emissions impacts of small, local changes in demand, since only a few specific generators are expected to change production in response. 
This response to marginal changes can be estimated both at the network level---quantifying the aggregate sensitivity across many nodes in the network---or at specific locations.  
Indeed, these metrics vary on a node-by-node basis, as network constraints and the local energy mix dictate which generators are available to compensate for changes in demand. 
Hence, \textit{locational} marginal emissions rates (LMEs)~\cite{Callaway2017LocationResources,Rudkevich2012LocationalMaking} quantify the emission sensitivity at the nodal level, revealing the spatial heterogeneity in marginal emissions rates that emerges from network constraints.
LMEs are the emissions-equivalent of locational marginal prices, which have been studied extensively in the power systems community due to their importance to electricity markets~\cite{Orfanogianni2007AEvaluation, Ji2017ProbabilisticCongestion, Bo2009ProbabilisticUncertainty}.
LMEs have been used to quantify the impacts of various policies on carbon emissions, e.g., increasing electric vehicle penetration~\cite{Tong2020WhatStates, Jochem2015Assessing2030} and changing electric vehicle charging policies~\cite{Pavic2015RoleSystems}, and are published live in the PJM Interconnection~\cite{FiveRates}, a major U.S.\ system operator.
LMEs have also been used in transmission expansion studies~\cite{Sun2017AnalysisAccounting, SaumaEnzo2018GlobalExpansion, Degleris2021Emissions-awareDifferentiation}. 
In this application, the LMEs define the marginal emissions effect of offsetting demand at a particular node and can be viewed as the gradient of emissions with respect to net load in the planning optimization problem~\cite{Degleris2021Emissions-awareDifferentiation}.

Empirical studies on marginal emissions rates in the U.S.\ and U.K.\ have used \textit{regression-based approaches} to estimate emissions rates across large geographical regions~\cite{Hawkes2010EstimatingSystems, Siler-Evans2012MarginalSystem, GraffZivin2014SpatialPolicies, Callaway2017LocationResources}.
These works leverage publicly available data and fit linear models to changes in emissions and net demand.
The main benefit of these methods is that they do not require a model of the underlying electricity system.
However, because of their inherent data requirements, these methods are difficult to extend to finer geographic resolutions and hypothetical electricity systems that lack preexisting data. 

In contrast, \textit{model-based approaches} explicitly calculate LMEs using the underlying dispatch model.
This calculation has been performed using marginal generator identification in merit-order models~\cite{Deetjen2019Reduced-OrderSector,Zheng2016AssessmentConstraints}, LMP-based approximations~\cite{Carter2011ModelingDelivery,Rogers2013EvaluationEmissions,
Wang2014LocationalDistribution
}, or, in simple cases, explicit formulae~\cite{Conejo2005LocationalSensitivities,Ruiz2010AnalysisNetworks, Rudkevich2012LocationalMaking}.
Model-based methods are promising because dispatch in real-world electricity systems often involves solving an optimization problem.
However, the models used in real world systems are highly complex, limiting the applicability of specific derivations in the aforementioned studies and highlighting the need for incorporating and feature specific constraints, e.g., ramping limits or storage, that need to be taken into account when calculating LMEs~\cite{Zheng2016AssessmentConstraints, Gil2007GeneralizedGeneration}.

For example, merit-order-based models usually neglect the impact of network constraints and only focus on generation cost to identify the marginal generator. 
LMP-based methods, which rely on matching LMPs with generation cost in order to identify the marginal generator, are not exact~\cite{Rogers2013EvaluationEmissions, Wang2014LocationalDistribution}, and the presence of complex coupling constraints would likely make identification even harder.
On the other hand, analytical derivations, while exact, have so far only been conducted on dispatch models with a limited number of constraint types, e.g., transmission constraints. 
As far as we are aware of, no previous studies have exactly accounted for time dependencies such as energy storage when computing LMEs.

In this work, we address this limitation by developing a method that supports arbitrary convex optimization-based dispatch models.

\subsection*{Contribution and outline}

This paper makes three main contributions:
\begin{itemize}
    \item 
    We propose \textit{a new method to compute LMEs} in arbitrary convex optimization-based dispatch models.
    This is in contrast to standard regression-based methods that may have significant data requirements and previous model-based methods that have been derived for specific dispatch models.
    The method we propose generalizes previous analytical derivations~\cite{Ruiz2010AnalysisNetworks,Rudkevich2012LocationalMaking} and is generic, flexible, and suitable for real systems dispatched by grid operators.

    \item
    We use the proposed method to \textit{derive LMEs in networks with dynamic constraints}, such as energy storage.
    As far as we are aware of, this is the first method to calculate LMEs for dynamic network dispatch models, such as the standard dynamic economic dispatch problem. 

    \item
    We \textit{demonstrate the utility of computing LMEs in dynamic models} using two different experimental setups. 
    First, using a published dataset~\cite{Deetjen2019Reduced-OrderSector}, we show
    that dynamic models more accurately represent the real-world relationship between demand and emissions compared to their static counterparts.
    Second, we use our method to study the impact of dynamic constraints on emissions rates using a realistic model of the Western United States transmission system. 
    The dynamic LMEs are distinct from their static counterparts, demonstrating the importance of accurately including all relevant dynamic constraints when estimating emissions sensitivities.
    
\end{itemize}

The paper is structured as follows. In Section~\ref{sec:problem}, we introduce the problem of computing LMEs in dynamic electricity networks.
We show that this problem generalizes previous approaches~\cite{Ruiz2010AnalysisNetworks, Rudkevich2012LocationalMaking} to complex models with temporal constraints.
We then show how to solve this problem using implicit differentiation in Section~\ref{sec:differentiation}.
Although we use this technique to compute LMEs for the model specified in Section~\ref{sec:problem}, our technique generalizes to arbitrary convex optimization-based dispatch models, including those used by system operators in real world electricity markets.
Lastly, we report simulation results on two datasets in Section~\ref{sec:experiments}. 
In the first experiment, we demonstrate the validity of our approach on real US electricity data and compare our results with an established method~\cite{Deetjen2019Reduced-OrderSector}.
In particular, we show that a dynamic model with unit-commitment constraints more accurately models changes in emissions compared to its static counterpart.
Second, we analyze a 240-bus model of the Western United States and show that, in the presence of grid-scale storage, computing LMEs dynamically is essential to accurately quantifying changes in emissions.
We conclude and discuss future work in Section~\ref{sec:conclusion}.

\section{Problem formulation}
\label{sec:problem}

In this section, we formulate the problem of computing the LMEs in a dynamically-constrained electricity system.
First, Section~\ref{sec:dispatch} provides background information on the \textit{dynamic dispatch problem}, a mathematical model for electricity networks with temporal constraints.
We then describe our mathematical model for emissions and marginal emissions in static networks in Section~\ref{sec:emissions}, where we formally state the dynamic marginal emissions problem.
Next, in Section~\ref{sec:special}, we describe two special cases of the dynamic marginal emissions problem that have been solved in previous work.
Notably, both special cases are static, i.e., they do not incorporate any dynamic constraints.
Finally, Section~\ref{sec:devices} gives three examples of dynamic devices, i.e., devices that cannot be represented in a static model.

\subsection{Dynamic dispatch model}
\label{sec:dispatch}

In electricity systems, a \textit{dispatch model} is a mathematical model for deciding which electricity generators should be used to meet demand.
Dispatch models are often formulated as convex optimization problems, where the variables are the amount of power produced by each generator, and the parameters include both the current electricity demand and the physical constraints on the system.
When modeling emissions, past work has often considered \textit{static} dispatch models, i.e., models that only reflect a single instant in time~\cite{Ruiz2010AnalysisNetworks, Rudkevich2012LocationalMaking, Li2013CarbonObligation, Deetjen2019Reduced-OrderSector}.
However, most real world electricity systems have \textit{dynamic} constraints---constraints that couple generator outputs across time periods.
For example, a generator with ramping constraints can only change its power output by a small amount between successive time periods.
In order to effectively model the impact of temporal constraints on emissions, we will study the \textit{dynamic optimal power flow problem},\footnote{%
The dynamic (DC) optimal power flow problem has been well studied in the power systems community~\cite{Xie2001DynamicMethods, He2019OptimizingApproach} and is sometimes referred to as the dynamic economic dispatch problem.
}
\begin{equation}
\label{eq:dyn-opf}
\begin{array}{lll}
    \textrm{minimize}
    & \sum_{j=1}^k f_j(g_j) \\[0.5em]
    \textrm{subject to} 
    & g_j \in \mathcal G_j, \quad & j \in [1:k], \\
    & \ones^T d_t  = \ones^T \tilde g_t, 
    & t \in [1:T], \\
    & F (d_t - B \tilde g_t) \leq u^{\max}, 
    & t \in [1:T],
\end{array}
\end{equation}
where the variable is $G \in \reals^{T \times k}$.
In the above, we use $g_j$ to refer to $j$-th column of $G$, and $\tilde g_t$ to refer to the $t$-th row of $G$.
The matrix $G$ represents the power output of $k$ devices over $T$ timesteps; each device can supply power, store power, or otherwise interact with the grid.
The entry $G_{tj}$ represent the output of device $j$ at time $t$.

\paragraph{Device Costs and Constraints}
Each device $j$ has three properties: a convex cost function $f_j(g) : \reals^T \rightarrow \reals$, a convex constraint set $\mathcal G_j \subset \reals^T$, and a location on the network.
We model the device locations with a matrix $B \in \{0, 1\}^{n \times k}$ that maps each device to a particular node in the network, i.e., $B_{i j} = 1$ if device $j$ is located at node $i$, and $B_{ij} = 0$ otherwise.
The objective of Problem~\eqref{eq:dyn-opf} is to minimize the sum of the device costs, and the first constraint states that each device must stay within their constraint set.

\paragraph{Network Constraints} 
Problem~\eqref{eq:dyn-opf} considers an electricity network with $n$ nodes and $m$ edges (transmission lines).
Node $i \in [1:n]$ has demand $(d_t)_{i}$ at time $t \in [1:T]$.
The second constraint is that power must be balanced across the network, i.e., $\ones^T d_t = \ones^T \tilde g_t$.
Finally, the third constraint is that the power flowing across each transmission line is limited by its capacity.
We define $F \in \reals^{n \times m}$ to be the \textit{power flow distribution factor matrix}, where $F_{i \ell}$ determines how a power injection at node $i$ (withdrawn at node $n$) affects power flow across line $\ell \in [1:m]$.
Because of thermal and voltage phase angle constraints, each line $\ell$ can only transport up to $u^{\max}_{\ell}$ units of power,
modeled with the constraints $F (d_t - B \tilde g_t) \leq u^{\max}$.

\paragraph{Solution Map}

Let $D = (d_1, \ldots, d_T) \in \reals^{Tn}$ be the concatenated vector of demand schedules and assume the solution to Problem~\eqref{eq:dyn-opf} exists and is unique for all $D \in \reals^{Tn}$.\footnote{%
We make this assumption with little loss of generality. 
For example, we can guarantee a solution exists by adding a generator with no capacity limits and extremely high costs to each node (representing curtailed demand).
Similarly, we can guarantee uniqueness by adding a small quadratic penalty $\sum_{t, j} G_{tj}^2s$ to the objective.}
Let $G^*(D) : \reals^{Tn} \rightarrow \reals^{T \times k}$ denote the optimal choice of $G$ in Problem~\eqref{eq:dyn-opf} as a function of $D$.
Because we assume the solution to Problem~\eqref{eq:dyn-opf} exists uniquely for all $D$, then $G^*$ is a well-defined function.
We call $G^*$ the \textit{solution map}, and use the vector-valued function $\tilde g_t^*(D) : \reals^{Tn} \rightarrow \reals^k$ to denote the $t$-th row of $G^*$.
As we will see shortly, the solution map will allow us to formalize the relationship between demand and emissions.

\subsection{Locational marginal emissions}
\label{sec:emissions}

We model the emissions of generator $i$ as a linear function of power output with rate $c_i$,
i.e., the total emissions at time $t$ are $c^T \tilde g_t$.
Since the generator power outputs are determined by the dispatch model, the emissions at time $t$ generated as a function of demand is $E_t(D) =  c^T \tilde g_t^*(D)$.
The \textit{total emissions} over the entire time horizon are then $E(D) = \sum_t E_t(D)$.
Although we use a linear model throughout the remainder of the paper, it is straightforward to generalize all our results to nonlinear models.
For example, if each generator has nonlinear emissions functions $\gamma_i(g) : \reals \rightarrow \reals$, then the total emissions at time $t$ is $E_t(D) = \sum_i \gamma_i( (\tilde g_t)_i )$.

\begin{figure}[t]
    \centering
    \includegraphics[width=3in]{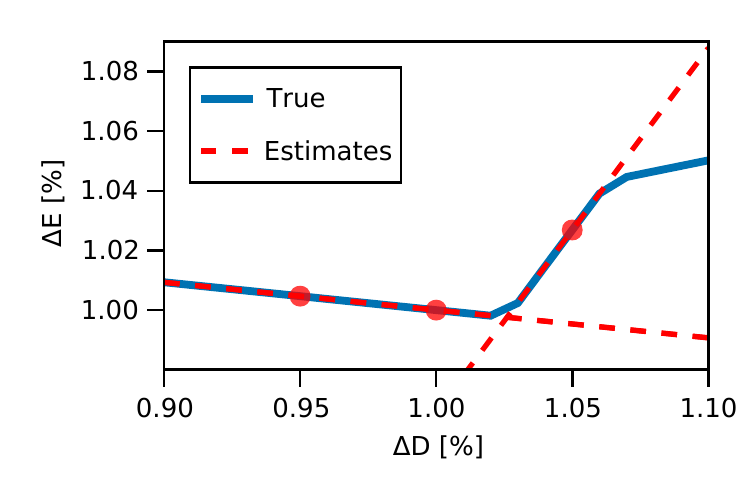}
    \caption{
    Illustration of marginal emissions rates.
    (Solid blue curve) Total emissions as a function of demand at a particular node.
    (Dashed red curves) 
    The first order approximations defined by the LMEs calculated at each red circle. 
    The LMEs are the slopes of the dashed red curves.
    }
    \label{fig:illustration}
\end{figure}

\paragraph{Problem statement}
The LMEs $\Lambda(D) : \reals^{Tn} \rightarrow \reals^{Tn}$ are the marginal rate of change in total emissions given a marginal change in demand at a specific node and at a given time.
In other words, the LMEs are the gradient of emissions with respect to demand, i.e., $\Lambda(D) = \nabla E(D)$.
The function $\Lambda(D)$ is vector-valued, since changes in electricity consumption at different nodes and different times may have different impacts on emissions.
As an illustration, we report a comparison between total emissions and LMEs for different values of demand at a given node in an arbitrary network (see Fig.~\ref{fig:illustration}). Locally, LMEs do indeed provide good approximations to the change in total emissions. It is however clear that those metrics are only locally valid and can sometimes be ill-defined, e.g. at points of non-differentiability of total emissions.
The problem we study in this paper is how to compute $\Lambda(D)$ when the solution maps $\tilde g_t^*(D)$ are determined by the dynamic optimal power flow problem. As far as we are aware, no prior published results have shown how to compute LMEs for generic dynamic dispatch models.

\subsection{Special Case: Static Generators}
\label{sec:special}

When we restrict the devices (the functions $f_i$ and sets $\mathcal G_i$) to be static generators, we recover previous analytical models~\cite{Conejo2005LocationalSensitivities,Ruiz2010AnalysisNetworks, Rudkevich2012LocationalMaking}.
The static generator device has constraint set $\mathcal G = \{ g \mid g^{\min} \leq g \leq g^{\max} \}$, where $g^{\min}, g^{\max} \in \reals^T$ and cost function $f(g) = \sum_t a g_t^2 + b g_t$.
The static generator could represent a traditional dispatchable generator, in which case $g^{\max} = \alpha \ones$, or a renewable resource, in which case the entries of $g^{\max}$ may vary.
Most importantly, the static generator has no temporal constraints: $g_t$ is independent of the choice of $g_{t'}$ when $t \neq t'$.
In a network with only static generator devices, the dynamic problem would simplify to $T$ \textit{static} optimal power flow problems that can be solved independently.
Moreover, if we remove the network constraints by setting $F = 0$, we recover the model used in~\cite{Deetjen2019Reduced-OrderSector} to empirically estimate emissions rates.

\subsection{Dynamic Devices}
\label{sec:devices}

By addressing the dynamic optimal power flow problem in its full generality, our framework allows us to consider dynamic devices as well.
These devices have temporal constraints, implying their actions at any given time depend on their actions at other times.
We give three examples below.

\paragraph{Ramp-constrained generators}
Ramping constraints limit the rate at which a generator can change its output.
These generators are modeled with the constraint set,
\begin{align*} 
\mathcal G = \left\{\right.  g \mid &   g^{\min} \leq g \leq g^{\max}, 
\\
& g_t - \rho \leq g_{t+1} \leq g_t + \rho, t \in [1:T-1]\left.\right\}
\end{align*}
where $\rho \in \reals^k$ is the ramping rate of each generator.
Ramp-constrained generator devices have the same cost functions as static generator devices, $f(g) = \sum_t a g_t^2 + b g_t$.
Ramping constraints are particularly useful in dynamic dispatch models with short time intervals, e.g., 15 minutes, and slow dispatching generators, like nuclear and hydro.

\paragraph{Storage devices}
\label{sec:battery}
Storage devices include pump-hydro resources, grid-scale batteries, and DER aggregators selling storage services to the grid. 
We define a storage device to have cost function $f(g) = 0$ and constraint set $\mathcal G$ that is the set of $g \in \reals^T$ such that there exists $s, \gamma, \delta \in \reals^T$ satisfying,
\begin{align*}
\arraycolsep=1.4pt
\begin{array}{rll}
    0 &\leq s \leq C, \quad\quad 
    0 \leq \gamma \leq P, \quad\quad
    0 \leq \delta \leq P, \\
    g_t &= \delta_t - \gamma_t, \quad\quad
    s_t = s_{t-1} + \eta \gamma_t - (1/\eta) \delta_t,
\end{array}
\end{align*}
for $t \in [1:T]$, where $s_0 = 0$.
The vector $C \in \reals$ is the storage capacity, $P \in \reals$ is the maximum (dis)charge rate, and $\eta \in (0, 1]$ is the storage efficiency.


\paragraph{Unit-commitment-constrained generators}
Unit-commitment constraints are integer constraints specifying whether or not a power generator is on.
If generator $i$ is on, it must produce a minimum power $g^{\min}_i$ and stay on for a specified period of time.
We model this by modifying the generator device constraint set to be the set $\mathcal G = \mathcal G_{\textrm{static}} \cup \{ 0 \}$, where $\mathcal  G_{\textrm{static}}$ is the equivalent static generator.
Although the set $\mathcal G$ is not convex, it can be made mixed-integer convex by introducing an integer variable $z \in \{0, 1\}$.
Although the mixed-integer constraint makes~\eqref{eq:dyn-opf} NP-hard, many good heuristics for solving mixed-integer convex programs are readily available through commercial solvers such as Gurobi~\cite{GurobiOptimization2022GurobiManual}.

\section{Implicit differentiation-based LMEs}
\label{sec:differentiation}

In previous model-based studies, locational marginal emissions rates are derived by first calculating how generation changes with demand, and then multiplying the change in generation by the emissions rate of each generator.
This is a manifestation of the chain rule,
which states that the LMEs are $\Lambda(D) = \nabla E(D) = \sum_{t=1}^T J \tilde g_t^* (D)^T c$,
where $J f(z) \in \reals^{k \times n}$ denotes the Jacobian of the function $f : \reals^n \rightarrow \reals^k$ evaluated at $z \in \reals^n$.
Therefore, the main technical challenge when computing $\Lambda(D)$ lies in computing an analytical expression for the Jacobians $J \tilde g_t^* (D)$.

In previous studies that only consider static dispatch models, i.e., $T = 1$, one only needs to derive a single expression for $J \tilde g_1^* (D) \in \reals^{k \times n}$. In the general setting, the situation is much more complex---one must derive $T$ Jacobians $J \tilde g_t^*(D)$ of size $k \times Tn$.
Although deriving an analytical expression might be possible, we take a simpler and more powerful approach in this paper: we use the \textit{implicit function theorem} to compute the Jacobians $J_D \tilde g_t^*(D)$. 
Our approach essentially generalizes the analytical derivations of~\cite{Ruiz2010AnalysisNetworks, Rudkevich2012LocationalMaking} to arbitrary convex optimization-based dispatch models, producing identical in the simpler static setting. 

\begin{theorem}[Implicit Function Theorem, \cite{Dontchev2014ImplicitAnalysis}]
\label{thm:implicit}
Suppose $K : \reals^n \times \reals^r \rightarrow \reals^k$ is strictly differentiable at $(D_0, x_0) \in \reals^n \times \reals^r$ and $K(D_0, x_0) = 0$.
Moreover, suppose $J_x K(D_0, x_0)$ is nonsingular, where $J_z f(z, y) \in \reals^{k \times n}$ denotes the partial Jacobian of the function $f : \reals^n \times \reals^r \rightarrow \reals^k$ with respect to $z$ evaluated at $(z, y)$.
Then the solution mapping $x^*(D) = \{ x \in \reals^r \mid K(D, x) = 0 \}$ is single-valued in a neighborhood around $(D_0, x_0)$ and strictly differentiable at $(D_0, x_0)$ with Jacobian 
$$
J x^* (D_0) = - J_x K(D_0, x_0)^{-1} J_D K(D_0, x_0).
$$
\end{theorem}

The implicit function theorem states that if a differentiable system of equations $K(D, x) = 0$ has a solution at $(D_0, x_0)$, and the corresponding partial Jacobian $J_x K(D_0, x_0)$ is non-singular, then the solution map $x^*(D)$ is a locally well-defined function with Jacobian given by Theorem~\ref{thm:implicit}.

In our setting, the solution map $G^*(D)$ is not the solution to a system of equations, but is rather the solution to a convex optimization problem. 
From convex analysis, we know that $G$ solves Problem~\ref{eq:dyn-opf} if and only if it solves the Karush-Kuhn-Tucker (KKT) equations $K(D, G) = 0$~\cite{Boyd2004ConvexOptimization, Dontchev2014ImplicitAnalysis}.
Therefore, we can apply the implicit function theorem to the KKT equations to compute $J G^*(D)$.
If we assume that the device objectives $f_j$ are twice differentiable and that the device constraints $\mathcal G_j$ can be parametrized via a set of twice differentiable inequalities, then the KKT equations are strictly differentiable, and $(D, G^*(D))$ satisfy the conditions of Theorem~\ref{thm:implicit}
(for more details on the implicit function theorem and its applications to optimization, we refer the reader to~\cite{Dontchev2014ImplicitAnalysis} and the references therein).
By combining this with the chain rule, we can then derive marginal emissions rates for the dynamic optimal power flow problem specified by Problem~\eqref{eq:dyn-opf}.
To summarize, we compute LMEs in two steps.
First, we compute the gradient of emissions with respect to the device outputs, i.e. $dE/d\tilde g_t = c$ in the case of linear emissions functions.
Then, we multiply this by the Jacobian $J G^*(D)$, which is computed using the implicit function theorem. The resulting metrics indicate the changes in emissions resulting from a marginal change in demand, as illustrated in Fig.~\ref{fig:illustration}.


Critically, this derivation works for any dynamic dispatch model that fits the form in Problem~\ref{eq:dyn-opf}, i.e., regardless of the choice of device cost functions $f_j$ and constraint sets $\mathcal G_j$ (as long as they are convex and twice differentiable).
When calculating the LMEs, different choices of device characteristics in Problem~\eqref{eq:dyn-opf} only change the KKT operator $K(D, G)$ and its corresponding partial Jacobians $J_G K(D, G)$ and $J_D K(D, G)$;
however, the general derivation using the implicit function theorem remains unchanged.
In practice, these Jacobians can either be derived analytically or computed using an automatic differentiation library, such as~\cite{Innes2018DontPrograms}, given a programmatic specification $f_j$ and $\mathcal G_j$.

\textit{Remark:}
Since the dynamic OPF problem includes the static OPF problem and the economic dispatch problem as special cases, this work generalizes the derivations in~\cite{Ruiz2010AnalysisNetworks, Rudkevich2012LocationalMaking} and~\cite{Deetjen2019Reduced-OrderSector}.
However, our method is by no means constrained to the dynamic dispatch model used in this paper;
implicit differentiation can be used to derive the marginal emissions rates for any convex-optimization based dispatch model.
Importantly, this includes many of the dispatch models used by system operators in practice, a point we revisit in Section~\ref{sec:conclusion}.

\subsection{Complexity and Software Implementation}
\label{sec:software}

We implement our method in Julia \cite{Bezanson2017Julia:Computing} for all the aforementioned dispatch models and constraints.
Our implementation is publicly available on GitHub in the package \verb_DynamicMarginalEmissions.jl_.
We use \verb_Convex.jl_~\cite{Udell2014ConvexJulia} to model the problem and solve it with external solvers, such as ECOS~\cite{Domahidi2013ECOS:Systems} or Gurobi~\cite{GurobiOptimization2022GurobiManual}.
For the large-scale network discussed in Sections~\ref{sec:nrel} and~\ref{sec:renew} (i.e., $n = 240$ nodes, $k = 136$ generators, $m = 448$ lines) and $T = 1$ time periods, our software package solves the dispatch problem and computes the resulting LMEs in just under a second on a modern laptop with a 2.3~GHz quad-core processor.
For the same problem with $T=24$ time periods, the same machine takes about two minutes to solve the problem and compute the LMEs. 

Our software package offers a flexible interface and can be used to analyze networks with different physical characteristics (e.g., locations of generators, transmission line ratings) and constraints (e.g., ramping rates). 
After specifying the network parameters, one can compute the LMEs with a single function call.
Because our implementation is open-source, reasonably fast, and easy to use, we believe it is of value to the broader community.

In general, we expect our method to scale well to realistic, large problems encountered by grid operators.
Specifically, let $z = T \cdot \max(m, n, k)$.
Solving the dispatch problem itself requires $O(z^4)$ operations. 
Once the dispatch problem is solved, constructing and inverting the Jacobian to compute LMEs only requires $O(z^3)$ operations, which is negligible compared to the complexity of solving the dispatch problem itself.
Since most grid operators must solve dispatch problems at regular intervals, e.g., every 15 minutes to clear the real-time market, computing LMEs can be viewed as a post-processing step requiring little additional compute time.



\section{Simulation Results} \label{sec:experiments}

In this section, we illustrate the applicability and utility of the suggested approach using two different simulation setups. First, in Section~\ref{sec:wecc-experiment}, we compute the LMEs for a static model and a dynamic model with unit-commitment constraints using real demand and generator data from the U.S.\ Western Interconnection~\cite{Deetjen2019Reduced-OrderSector}.
We compare each model's LMEs to real changes in emissions and to estimates from a merit-order-based model~\cite{Deetjen2019Reduced-OrderSector}.
Second,
in Sections~\ref{sec:nrel}, ~\ref{sec:renew} and \ref{sec:static_vs_dynamic},
we illustrate the methodology on a recent reduced network model of the same region~\cite{Yuan2020DevelopingIntegration}. 
Using the original dataset, we highlight the geographic variance of marginal emissions across the network. 
Then, we investigate the potential impacts of hypothetically large renewable penetration of storage and renewable generation. 
We conclude by comparing the LMEs obtained from a static approximation to those of the dynamic model.

\subsection{Economic Dispatch in the Western United States}
\label{sec:wecc-experiment}

In our first experiment, we analyze electricity data from the U.S.\ Western Interconnection system in 2016.
The Western Interconnection dataset is compiled in \cite{Deetjen2019Reduced-OrderSector} and contains weekly generator capacities, generator costs, and generator emission rates for large (above 25 MW capacity) fossil fuel generators in the Western Interconnection, as well as hourly total demand and total carbon emissions.
Because no transmission data is available, we consider models without transmission constraints.
The LMEs for a static model, a dynamic model with unit commitment constraints, and a state-of-the-art static merit-order method are compared to the real hourly rate of change in emissions.

\paragraph{Models}
We analyze two models, which we compare to a baseline. 
First, we analyze the results of the simple economic dispatch model \eqref{eq:dyn-opf}, with linear costs $f_i(g_i) = b_i g_i$, where $b_i$ is the marginal operation cost of generator $i$.
Second, we analyze a dynamic economic dispatch model with unit commitment constraints, over a time horizon of $T = 24$.
The unit-commitment constraints are only applied to coal generators, all of which are given a minimum output of $g^{\min}_i = 0.4 g^{\max}_i$.
We benchmark our results against the \textit{reduced-order dispatch model (RODM)} described in~\cite{Deetjen2019Reduced-OrderSector}.
The core of the RODM is a \textit{merit order}-based dispatch process: generators are dispatched in ascending order of cost.
After dispatching generators via the merit-order, the \textit{marginal generator}---the generator next in line to modify its output to meet an increase in demand---is identified to find the marginal emissions rate of the system.
In~\cite{Deetjen2019Reduced-OrderSector}, post-processing steps are applied to generate the marginal emission rates.
Notably, when no post-processing is applied, the RODM is identical to the economic dispatch model in \eqref{eq:dyn-opf} with linear costs $f_i(g_i) = b_i g_i$.

\paragraph{Results}
After generating LMEs $\lambda_t$ for every hour $t = 1, \ldots, T$ of the year, where $T = 8760$, we compare the resulting LMEs to the actual hourly changes in emissions.
Specifically, we compute the 
change in demand $\Delta d_t$ and change in emissions $\Delta E_t$ for every hour of the year.
Each model's estimated change in emissions is given by $\Delta \hat{E}_t = \lambda_t \Delta d_t$.
In order to compare the 
models, we compute the absolute error
$|\Delta \hat{E}_t - \Delta E_t| / Z$ 
of each model's estimate against the true hourly change in emissions at each timepoint, where errors are normalized by the mean absolute change in emissions $Z = (1/T) \sum_{t=1}^T |\Delta E_t|$.
We use absolute error, instead of square error, to minimize the effect of outliers.

\begin{figure*}
\centering
\includegraphics[width=6.5in]{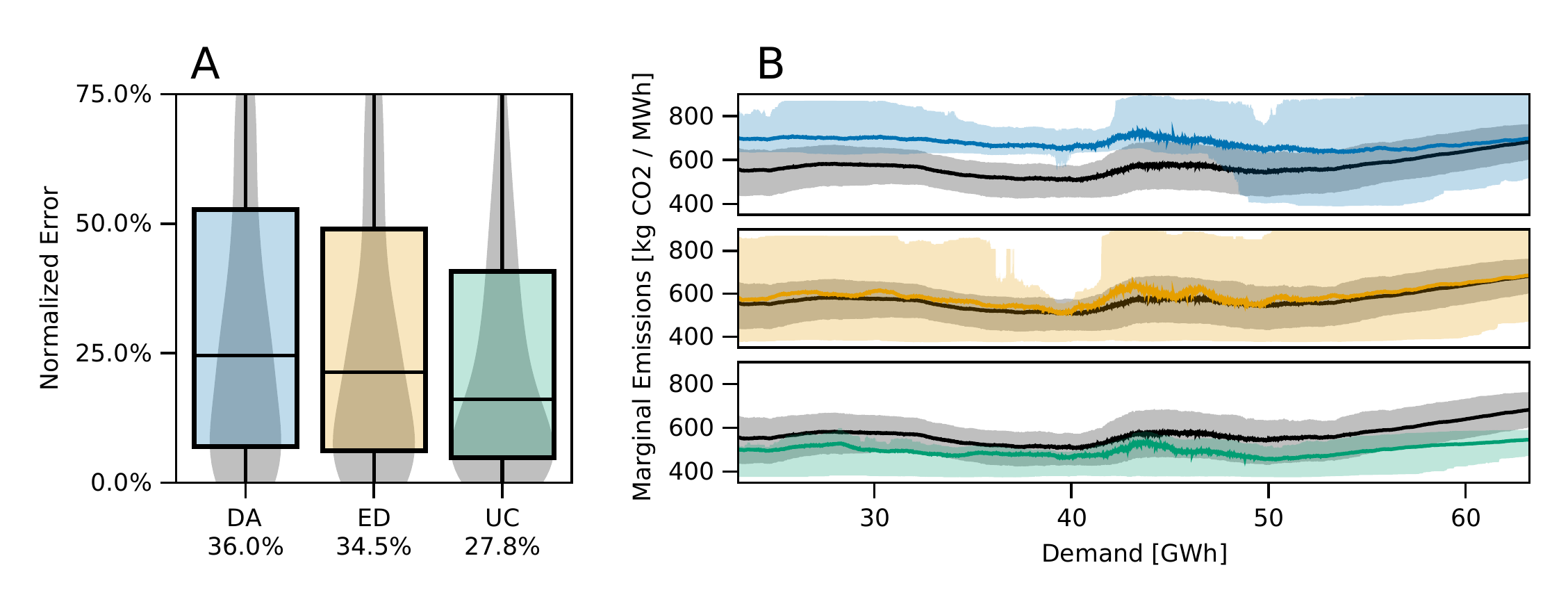}
\caption{
(Panel A) Emissions error for model from \cite{Deetjen2019Reduced-OrderSector} (DA), economic dispatch model (ED), and unit-commitment model (UC), normalized by the mean absolute change in emissions.
The ED model, effectively the same as the DA model without post-processing, performs similarly as DA. 
The UC model reduces the error by 8.2\% compared to the DA model, suggesting unit-commitment more accurately represents the real world dispatch process.
(Panel B) LMEs as a function of total demand.
Emissions rates are smoothed using the mean of a rolling window of 20\% of the data.
Shaded regions represent the interquartile range (IQR), i.e., the middle 50\% of the data, of the rolling window.
}
\label{fig:wecc}
\end{figure*}

A violin plot of absolute errors is displayed in Figure~\ref{fig:wecc}, Panel~A.
As expected, the economic dispatch model and the merit-order model from \cite{Deetjen2019Reduced-OrderSector} perform similarly---the merit-order model only differs from the economic dispatch model in its post-processing.
Notably, the unit-commitment model 
better models hourly changes in emissions than
both the economic dispatch model and the merit-order model,
reducing the mean absolute error by 8.2\%.
We attribute this to the fact that the unit-commitment model accurately represents dynamic constraints that appear in real-world dispatch processes, namely that coal generators cannot rapidly turn on and off again.

LMEs as a function of demand are also reported in Panel~B of Figure~\ref{fig:wecc}.
Historical LMEs are computed as $\lambda_t = \Delta E_t / (\Delta d_t + \epsilon)$, where $\epsilon = 0.5\ \text{MWh}$ is a small value used to avoid unreasonably large LMEs when $\Delta d_t$ is small.
Following a similar procedure to \cite{Deetjen2019Reduced-OrderSector}, the LMEs for the data and for each model are smoothed using the mean of a rolling window of 20\% of the data.
Shaded regions representing the interquartile range (IQR) of the data are also plotted to better understand the variance of each model.
After averaging, the LMEs produced by the economic dispatch model most closely resemble the data.
However, the variation is significantly reduced in the unit-commitment model, and the IQR most closely resembles that of the data compared to both other models. 

\subsection{240-bus Network Model}
\label{sec:nrel}

In this experiment, we study LMEs using a recent 240-bus network that models the Western United States in 2018~\cite{Yuan2020DevelopingIntegration}.
The dataset includes generator capacities, costs, and ramping rates; hourly demand and renewable generation for $T = 8760$ hours; and the efficiencies and capacities of four pumped hydro storage units.
We solve the dispatch problem and compute hourly LMEs using a dynamic model with storage devices as described in Section~\ref{sec:problem}.
We use this experiment to demonstrate the impact of network constraints on the LMEs.
Since the network has relatively little storage and only a few generators with ramping constraints, we do not analyze the impact of storage and dynamic constraints in this experiment.

\begin{figure*}
    \centering \includegraphics[width=6.5in]{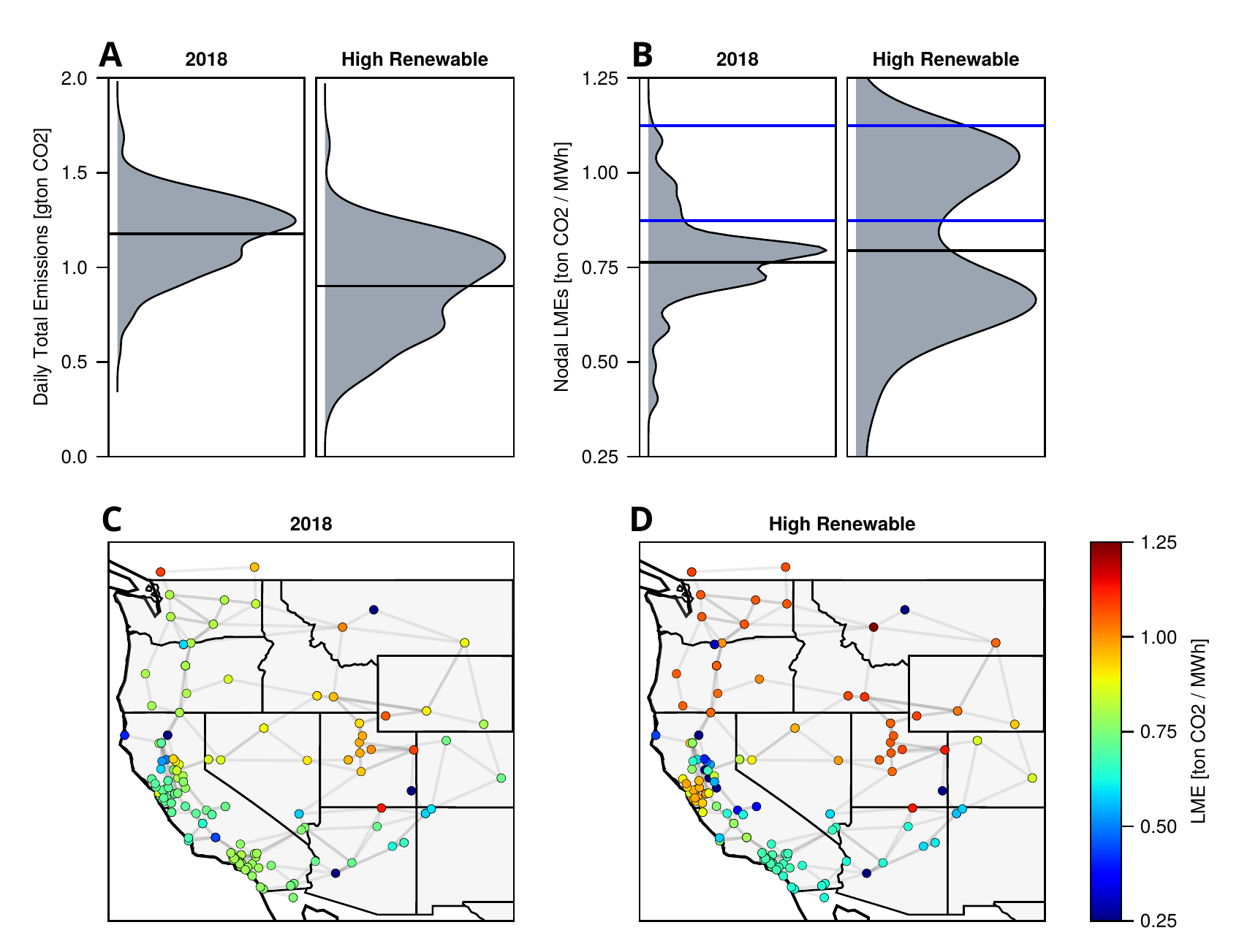}
    \caption{
    (Panel~A) Distribution (over days of the year) of network-wide, total daily emissions for both the 2018 case and the high renewable scenario.
    The mean of the distribution is denoted with a horizontal black line.
    (Panel~B) Distribution (over nodes and days of the month) of LMEs during August at 6pm, both for the 2018 case and the high renewable scenario.
    The mean of the distribution is again denoted with a horizontal black line.
    (Panel~C) A map of nodal LMEs through the 240-bus WECC network during the same time period (averaged over the month).
    (Panel~D) Same as Panel~C, but for the high renewable scenario with 15~GW of 4-hour storage.
    }
    \label{fig:240-map}
\end{figure*}

We report the distribution of daily total emissions in Figure~\ref{fig:240-map}, Panel~A (left frame) and the distribution of LMEs at 6pm in August (left frame) in Panel~B.
We observe that, on average, the distribution of nodal LMEs is narrowly concentrated around its mode.
However, we also note that the distribution of LMEs has relatively long tails;
although most of the LMEs are close to the mode, a few have drastically different values.
In Panel~C, we display a map of LMEs at 6pm in August (averaged over days of the month), illustrating the large geographic variation in emissions rates.
Panel~C demonstrates how transmission constraints can create large discrepancies in LMEs even within small geographic regions.
For example, despite their geographic proximity, the California Bay Area differs significantly from the Sacramento area.
The local diversity of the LMEs emphasizes the importance of modeling the network when computing emissions rates: a small geographic change can cause large differences in emissions rates. 

\subsection{High Renewable Scenario in the 240-bus Network}
\label{sec:renew}

To illustrate the impact of grid-level changes on emissions, a high-renewable version of the 240-bus network is presented in this section. 
Specifically, we uniformly scale renewable generation in the original 2018 model so that renewable generators meet 27\% of total demand (compared to 13.5\% originally).
We also add 15~GW of 4-hour battery storage
distributed among the ten largest renewable nodes proportional to generation capacity.
These batteries have a round trip efficiency of $89.8\%$ with symmetric charging and discharging efficiencies and are constrained to start and end each day with $50\%$ of their total capacity.
As in Section~\ref{sec:problem}, we assume the grid operator manages these batteries to minimize the overall cost of electricity.
%
The right frame of Panels~A and~B in Figure~\ref{fig:240-map} show the distribution of total emissions and LMEs in an identical manner to the 2018 case.
Similarly, Panel~D displays a map of average nodal LMEs akin to Panel~C. 
The 2018 and the high renewable scenarios differ in several ways.
First, as expected, \textit{total} emissions decrease significantly in the high renewable scenario. 
The changes to the locational \textit{marginal} emissions rates, on the other hand, vary significantly from node to node.
For example, adding renewable generation and storage causes LMEs to decrease at nodes in southern California, but to increase at nodes in Oregon and Washington.
In general, the distribution of nodal LMEs exhibits high variance and displays two modes, in contrast with the 2018 case.\footnote{We observe these differences for most hours of the day, but only display results for 6pm for concision.}
Overall, the changes in LMEs are complex and unintuitive: because of the presence of storage, nodal LMEs depend not only on non-local generation and transmission constraints, but also on that of every other time period. We believe this complexity is one reason grid operators should use an exact method for calculating LMEs (instead of relying, for example, on network-wide heuristics).





\subsection{Comparison Between Static and Dynamic LMEs}\label{sec:static_vs_dynamic}

In order to demonstrate the value of explicitly integrating dynamic constraints, we compare the true dynamic LMEs to the analogous ``static LMEs'' that arise from eliminating dynamic constraints.
Specifically, we consider how the LMEs would differ between static and dynamic models with the exact same loads and device outputs. 
Since static models cannot incorporate dynamic devices, we first solve the dynamic dispatch problem, then fix the battery charging and discharging schedules and consider them as parameters to a series of independent static problems, eliminating any dynamic constraints between subsequent timesteps.
We then compute the LMEs of the resulting model, which now only has static devices and constraints.

More formally, consider the dynamic optimal power flow problem in~\eqref{eq:dyn-opf}, with the devices ordered so that the first $k_1$ devices are static and the remaining $k_2$ devices are dynamic.
After solving the dynamic problem with $k = k_1 + k_2$ devices to obtain device outputs $ G^{*}$ and LMEs $\Lambda^*$, we solve the static problem,
\begin{equation}
\label{eq:static-opf}
\begin{array}{lll}
    \textrm{minimize}
    & \sum_{j=1}^{k_1} f_j(g_j) \\[0.5em]
    \textrm{subject to} 
    & g_j \in \mathcal G_j, \quad & j \in [1:k_1], \\
    & g_j =  g_j^*, \quad & j \in [k_1+1:k], \\
    & \ones^T d_t  = \ones^T \tilde g_t, 
    & t \in [1:T], \\
    & F (d_t - B \tilde g_t) \leq u^{\max}, 
    & t \in [1:T],
\end{array}
\end{equation}
where the variable is again $G \in \reals^{T \times k}$.
Since the schedules of the $k_2$ dynamic devices are fixed, $\tilde g_t$ is independent of $\tilde g_{t'}$ for $t \neq t'$, and Problem~\eqref{eq:static-opf} can be decomposed into $T$ independent optimization problems, if desired.
We compute the resulting `static' LMEs $\Lambda^{\textrm{static}}$ from solving Problem~\eqref{eq:static-opf}, and compare them to $\Lambda^*$.
In theory, the difference between the LMEs of a dynamic model and its static approximation can be arbitrarily large, as seen in the following example.
\begin{figure*}[!t]
    \centering
    \includegraphics[width=6.5in]{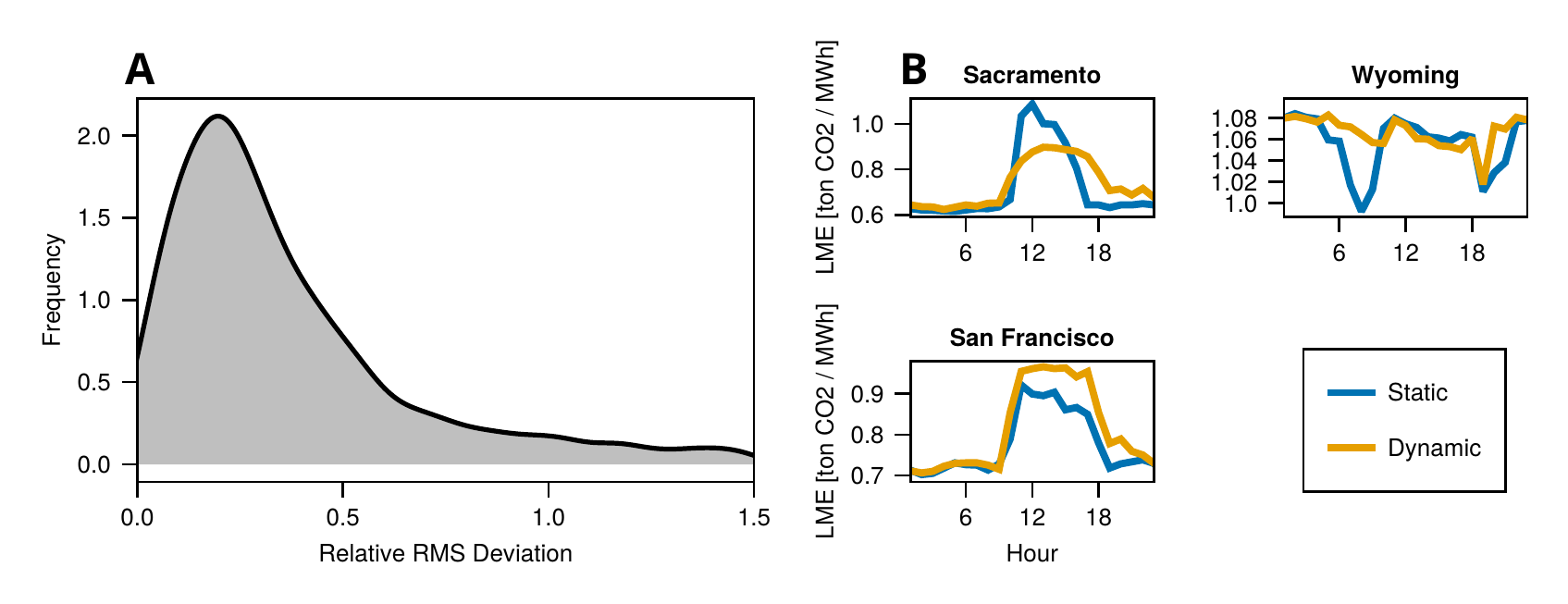}
    \caption{
    (Panel A) Distribution of root mean squared (RMS) deviation between nodal LMEs produced by the static model and the dynamic model, normalized by the median LME.
    The average RMS deviation between the static and dynamic model is 28.40\%.
    (Panel B)
    Hourly time series of static LMEs (blue) and dynamic LMEs (yellow) during three sample days.
    }
    \label{fig:240-error}
\end{figure*}

\textit{Example:} 
Consider a single node network with $T = 2$ timesteps and $k=3$ devices.
The first device is a gas generator with constraint set $\mathcal G_1 = [0, 10] \times [0, 10]$ (i.e., the generator has capacity 10 in both time periods), cost $f_1(g) = 1$, and emissions rate $c_1 = 500$.
The second device is a solar panel with constraint set $\mathcal G_2 = [0, 10] \times \{0 \}$ (i.e., the generator has capacity 10 in the first period, and no capacity in the second period), cost $f_3(g) = 0.1$, and emissions rate $c_3 = 0$.
Finally, the third device is a battery with constraint set $\mathcal G_3$ specified by Section~\ref{sec:battery}, with capacity $C = 10$, charging rate $P = 10$, and efficiency $\eta = 1$.
Assume a constant demand schedule $d = (1, 1)$.
The economic dispatch will result in the following device outputs: $g_1 = (0, 0)$, $g_2 = (2, 0)$, and $g_3 = (-1, 1)$, i.e., the solar panel will produce two units of power, storing some of it in the battery to serve the second time period, and curtail the remaining 8 units.
The dynamic LMEs are naturally $\Lambda(D) = (0, 0)$: if we had slightly higher demand in either period, the solar panel would curtail its output less to meet demand.
Now, if we fix the battery charging schedule to $g_3 = (-1, 1)$, solving the static problem gives the same device outputs $g_1 = (0, 0)$ and $g_2 = (2, 0)$.
However, the resulting static LMEs are $\Lambda(D) = (0, 500)$, a drastically different result.
This is because the static approximation fixes the battery schedule, so changes in demand during period two are met by a change in the gas generator output.

The toy example above demonstrates that dynamic and static LMEs can differ significantly in theory.
We verify that this occurs in practice using the 240-bus network from Section~\ref{sec:static_vs_dynamic}, where we use the same procedure to compute dynamic LMEs and their static approximations.

We report these differences in Figure~\ref{fig:240-error}, Panel A, where we display the distribution across all nodes and all days of the year of the root mean squared (RMS) deviation between the vector of daily emissions rates for the static model, $\lambda_{\textrm{static}} \in \reals^{24}$, and the dynamic model $\lambda_{\textrm{dynamic}} \in \reals^{24}$.
The average RMS deviation (normalized by the median LME) is 28.40\%, indicating that the static and dynamic models yield significantly different results.
In Panel B, we illustrate the static and dynamic LMEs for three randomly sampled days.
While static LMEs are very good approximations in some instances (e.g., morning hours in top-left), they deviate significantly in others.
These results suggest that ignoring dynamic constraints and simply computing static LMEs is not sufficient to model emissions in dynamic networks: explicitly computing dynamic LMEs is essential to understanding emissions rates in systems with significant dynamic constraints, such as large grid storage capacity.

\section{Conclusion} \label{sec:conclusion}

In this paper, we introduce a novel method for computing locational marginal emissions rates using implicit differentiation.
We use this method to compute the LMEs of dynamic dispatch models, i.e., dispatch problems containing temporal constraints. 

Using real WECC electricity and emissions data, we find that incorporating these dynamic constraints improves model accuracy by 8.2\%.
Finally, we observe that dynamic LMEs are difficult to approximate with their static counterparts: in a synthetic approximation of the WECC network, static and dynamic LMEs have a normalized average RMS deviation of 28.40\%.
Since flexible loads and energy storage are expected to play a large role in future grids, we believe incorporating dynamic constraints will be essential to accurately modeling LMEs.

The method presented in this paper generalizes previous methods to arbitrary convex optimization-based dispatch models.
Since many system operators use convex optimization-based dispatch models in day-ahead and real-time electricity markets \cite{PJM2021PJMOperations}, they could use this method to publish marginal emissions factors in real time.
Although these models can be notably more complex than those analyzed in academic research, the proposed method can compute marginal emissions factors for any such model, as long as they can be represented as convex optimization programs.
Moreover, by leveraging automatic differentiation software and optimization modeling languages~\cite{Dunning2017JuMP:Optimization}, the system operator would only need to specify the objective and constraints of their dispatch problem.
LMEs could then be published alongside LMPs to provide real time emissions information to electricity market participants.
This could be helpful, for example, to a large internet company choosing to reduce emissions by directing internet traffic to servers in low emitting regions, a problem considered in \cite{Lindberg2021AShifting}, or more generally to operators wanting to define optimal load management strategies~\cite{Wang2014LocationalDistribution}.

Finally, we comment on three directions for future work.
First, our experimental results indicate that LMEs in dynamic models often display complex behaviors and are difficult to interpret due to temporal and network constraints. 
Deciphering the mechanisms underlying the structure of the LMEs in different settings would be useful in communicating these results  and translating them into grid planning or policy decisions.
Second, we note that computing LMEs in large networks could be computationally intensive.
Exploiting network structure and using distributed computation could yield significant performance gains.
Third, our paper shows how to compute LMEs when the full network model is available.
In some cases, however, the network model may be unavailable to the interested party.
Understanding how to estimate the parameters of the electricity network from publicly available data (using the methods developed in \cite{Donti2018InverseData}, for example) and then deriving marginal emissions factors from the learned model is an interesting area of research.


\section*{Acknowledgements}

The authors thank Liang Min and In\^{e}s Azevedo for their valuable comments and suggestions.

Disclaimer: This report was prepared as an account of work sponsored by an agency of the United States Government. Neither the United States Government nor any agency thereof, nor any of their employees, makes any warranty, express or implied, or assumes any legal liability or responsibility for the accuracy, completeness, or usefulness of any information, apparatus, product, or process disclosed, or represents that its use would not infringe privately owned rights. Reference herein to any specific commercial product, process, or service by trade name, trademark, manufacturer, or otherwise does not necessarily constitute or imply its endorsement, recommendation, or favoring by the United States Government or any agency thereof. The views and opinions of authors expressed herein do not necessarily state or reflect those of the United States Government or any agency thereof.

\bibliographystyle{ieeetr}
\bibliography{references_edited}

\end{document}